# ATOMIC FORCE MICROSCOPY WEAR CHARACTERIZATION FOR METALLIC STENT POLYMER COATINGS

Luigi Lazzeri[1], Maria Grazia Cascone[1], Piero Narducci[1], Nicola Vitiello[1], Mario D'Acunto[1], Paolo Giusti[1,2]

[1]Department of Chemical Engineering, University of Pisa, 56126, Pisa, Italy, e-mail: m.dacunto@ing.unipi.it
[2]CNR, Institute of Composite and Biocompatible Materials, 56126, Pisa, Italy

**ABSTRACT**
Atomic force microscopy (AFM) has become established as a powerful and a versatile tool for investigating local mechanical properties. In addition, it has been made possible to take advantage of the AFM tip-sample interaction, to perturb, and in turn, to modify the surface of soft samples, such as polymers. The accurate knowledge of their response to the continuous AFM scanning could help to design new materials having desirable mechanical properties. In this paper, we present the results obtained applying a new methodology to investigate wear properties on two different type of polymers, such as poly(methyl-methacrylate) (PMMA), and poly(L-lactic acid) (PLLA). These polymers have been widely employed in biomedical applications and have recently been considered as good candidates for coronary metallic stent coatings.

**INTRODUCTION**
A fundamental understanding of surface properties of biomaterials on nanoscale scale should be generated in order to have a well satisfactory knowledge of responses of the tissue to biomaterials thereby minimizing or eliminating tissue trauma on macrometer scale [1]. There has been a substantial increase in the use of biodegradable or biostable polymers in the last decade. Initially, they were employed as carriers for drug-delivery devices, sutures and as a temporary joint spacer. However, the tribological characteristics of these polymers have not been fully investigated. Two aspects can limit the use or efficiency of such polymers: chemical degradation and/or wear. Chemical degradation starts breaking the long polymer chain into smaller fragments, or chain scission. Or, some enzymes and ionizing radiation are also capable of attacking polymers. Thus, the polymer may be reduced in molecular weight, increasing its solubility or it may become harder and more brittle due to cross-linking. Relative motion between parts can cause mechanical damage and release of small debris due to wear. The surface of an implanted polymer, such as any other material, is not perfectly smooth on a microscopic scale but rather has small asperity on the surface. Mechanical contact is localised to the asperity. Thus, a relatively low contact pressure for the entire surface can result in very high local pressures relatively to any single asperity. Such localised contact pressures can result in adhesion between asperity of two relatively motion surfaces. After adhesion, subsequent movement can provoke the formation of debris or small fragments. These fragments may react with other chains to form side branches, or react with other chain to form cross-links.
The aim of the present paper is to investigate the wear properties on micro/nanoscale making use of an AFM for PMMA and PLLA. PMMA is an amorphous material with good resistance to dilute alkalis and other organic solution. It is known for its exceptional light transparency, wear resistance and biocompatibility. PMMA is used broadly in medical applications such as blood pumps and reservoirs, membranes for blood dialyser. PLLA is a biodegradable polymer widely employed in tissue regeneration of cartilage, bone, skin, ligament, bladder and liver. These polymers could become two good candidates for coronary metallic stent coatings. It is well known that the implantation of metallic stents is a widely accepted alternatively therapy to percutaneous transluminal coronary angioplasty [2]. The application of stents shows promising results to prevent acute restenosis. In spite of this advantage, sub-acute restenosis occurs frequently despite of a high antithrombotic pharmaceutical regime. This is meanly due to the thrombogenicity of metals, which may be explained by their positive net electrical surface potential. An interesting approach to overcome those limitations is to cover the metal surface with a bioactive surface. This approach corresponds with a recent trend in biomaterials research combining desirable mechanical properties of an existing material with improved biocompatibility of another. Some efforts have been made to coat metallic stents with polymers in order to improve their blood compatibility. PLLA and PMMA could be good candidates to coat metallic stents rather than traditional polyamide, polyurethane or silicone.

**CHARACTERIZATION OF WEAR FOR PLLA AND PMMA ON MICRO/NANOSCALE**
There are different basic factors in providing the efficient wear test, such as type of motion, type of loading, lubrication specimen preparation and environment control. Investigation of the fundamental characteristics of wear at the micro-scale is complicated by few factors that are not critically addressed in the tribology of macro-systems. Since these forces are sensitive to the environment and surface condition of the specimens, it is quite difficult to determine the forces accurately. Further, quantification of wear is not straightforward since the amount of wear is often too small to be detected by surface-sensitive instruments. Micro-wear measurements have been object of fast development of precise measuring tools followed to the introduction of AFM [3-8]. The knowledge of wear mechanisms on micro/nanoscale could help in order to quantify the distribution of material loss during relative motion surfaces. In fact, weighing the sample before and after the test has been the dominant wear quantification technique. A precision balance typically has a resolution of $10^{-6}$ of the maximum load, which puts a limit to the minimum load possible to be quantified in relation to the total weight of the component. Moreover, the mass of polymers can change due to water absorption. Further, there is the remarkable point that on harder materials macro-wear is 10-1000 times higher than micro-wear, even though the mean contact pressure in the micro-test is higher than in the macrotest. In fundamental research, studies of the temporal evolution of tribological systems starting at the first few stressing cycles achieve growing interest. The smallest wear scars, often only some tens of micrometers in size, require a quantitative technique with a nanometer scale resolution.
Formation of debris by biomedical polymers could have serious consequences for human health. Coronary stents are exposed to two main stresses, first one is due to the contact with blood vessel and the second one is the shear stress provoked by the blood flow. Detachment of small wear debris from the polymer coating of metallic stent could stimulate thrombogenesys. This mechanism is not completely understood, for this reason, such as biocompatibility, haemocompatibility and biodegradation, wear characterization of biomedical polymers could become one important test to define suitable candidates for metallic stent polymer coatings.
The experimental and numerical methods adopted in this paper give the possibility to have good indications on the incidence of different wear-mechanisms. In turn, they should permit a good degree of comparison to be made of the various polymers employed in biomedical applications.

**MATERIALS AND METHODS**
**Sample preparation**
The PMMA (Depart. of Chemical Engineering, Pisa) and PLLA (Boheringer Ingelheim) used had molecular weight $M_w$ ca. 366000 and $M_w$ ca.380000, respectively. The structure of the monomers are $[-CH_2C(CH_3)(CO_2CH_3)-]_n$ for the PMMA and $[-OCH(CH3)CO-]_n$ for the PLLA. Two different solutions of the polymers in chloroform (Carlo Erba Reagenti) at 1% w/v and 0.25% w/v was prepared and deposed on a solid substrate of *SiO2* in air at temperature of 20 °C and relative humidity of 40%. Young's modulus and Poisson number for the materials involved during the scratching tests are reported in table 1.

| material | Young modulus (Gpa) | Poisson number ν |
|---|---|---|
| PMMA | 3 | 0.35 |
| PLLA | 0.8 | 0.41 |
| *SiO₂* | 70 | 0.17 |
| Silicon tip | 129 | 0.28 |
| Silicon nitride tip | 260 | 0.25 |

Table 1. Mechanical properties of samples and AFM probe tip

In some experiments, the sample area was formed by two hardness coefficient materials such as the polymer and *SiO₂*. Due to these different hardness coefficients, the response of the two materials to the stress induced by the repeated passage of the AFM probe tip was significantly different. The area of *SiO₂* not covered by the polymer can be considered a zero wear reference.

**Biomedical polymer wear tests**
The technique for wear testing consisted in using the AFM probing tip to abrade the surface of interest while simultaneously imaging the area where the polymer was being progressively damaged by the scanning tip. This technique permitted the following wear properties to be observed both qualitatively and, when possible, quantitatively: i) formation of ripples on the surface of the polymer; ii) qualitatively evolution of the surface before and after the test; iii) evaluate wear volume; iv) study the adhesion effect and subsequent degradation of probe tip. A δ-Silicon wafer grating (TGT1, NT-MDT, Russia) was used to test the tip degradation.
The measurements were carried out making use of an Autoprobe CP AFM (Park Scientific Instruments, Sunnyvale, CA) operating in contact mode with normal force varying in the range 1.0÷3.0nN. Images of different areas (256pixels×256 pixels) were acquired both with a silicon nitride tip (pyramidal shape, nominal probe radius 40nm, cantilever stiffness 0.03N/m) and with a silicon tip (conical shape, nominal probe radius of 10nm, cantilever stiffness 0.24N/m). The data were treated both with the image processing software from Park Scientific Instruments and with a MATLAB code written *ad hoc* to manipulate AFM data [9]. Quantitative analysis was carried out in some cases on the data as acquired, in other ones the background slope of the images was removed by a flattening routine. A best seventh-order polynomial fit to the average height profile was subtracted from each line of the entire image to take account of possible distorsions due to lateral displacements of the sample surface below the AFM tip.

**RESULTS AND DISCUSSION**
The mechanical contact between the probe tip and the polymer surface is defined by the following parameters: real area of contact $A$, penetration depth of the probe tip $z$, and yield stress $\tau$. These parameters can be approximated by a model which mix the Hertzian model of elastic contact of a sphere and a flat surface with other models accounting adhesion force contribution and possible high applied load plastic deformation [10,11]

$$A = \pi \left( \frac{3R_t F}{4E^*} \right)^{2/3} \qquad z = A/2\pi R_t \qquad \tau = F_n/A \qquad (1)$$

where $R_t$ is the tip radius of curvature and $E^*$ is the reduced Young's modulus of tip and polymer. The values of reduced Young's modulus are $E^*_{PMMA/silicon}=3.3378$; $E^*_{PLLA/silicon}=0.9551$. Equivalent results are obtained with the mechanical valued of silicon nitride. Some significant values of the real area of contact are $A_{PMMA/Silicon}$(applied load=3nN)≈11.2×10⁻¹⁸ m², $A_{PLLA/Silicon\ nitride}$(applied

load =1nN)≈31.24×10$^{-18}$m$^2$. The depth of penetration values vary in the range 0.02÷0.5nm, and yield stress values vary in the range 32÷ 268 Mpa.

One relevant aspect in AFM study of polymer wear on nanoscale is the formation of ripples. The ripples can be considered a consequence of elastic instability waves. They have been observed on macroscopic length scales for elastically soft materials such as rubber during sliding on hard surfaces. They were produced making use of high load regime contact mode, 100nN, and then observed on high Young modulus polymeric film [12]. The formation of ripples is commonly considered to be the result of a peeling process operated by the microscope tip on the polymer.

In order to demonstrate this process on our samples, an area completely covered by PMMA has been scanned for 30 cycles at 3nN with conical shape probe tip. Figure 1 shows the formation of ripples. Although the first cycle surface is not completely flat, nevertheless, the change of topography is quite evident after 30 cycles. Same tests on PLLA showed a drastic change in tip performance due to degradation both of the sample and probe tip. This result is important because it provides confirmation that ripple formation on nanoscale is a gift of high Young's modulus polymers.

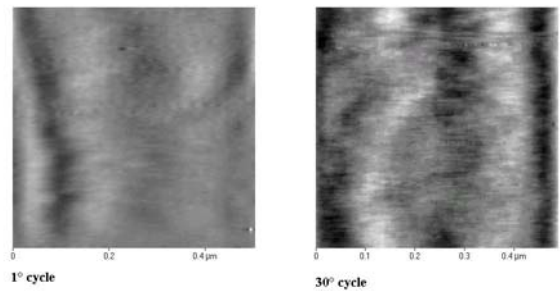

Figure 1. Evidence of ripples for PMMA after 30 probe tip scans (applied load 3nN, conical shape tip with nominal radius 10nm).

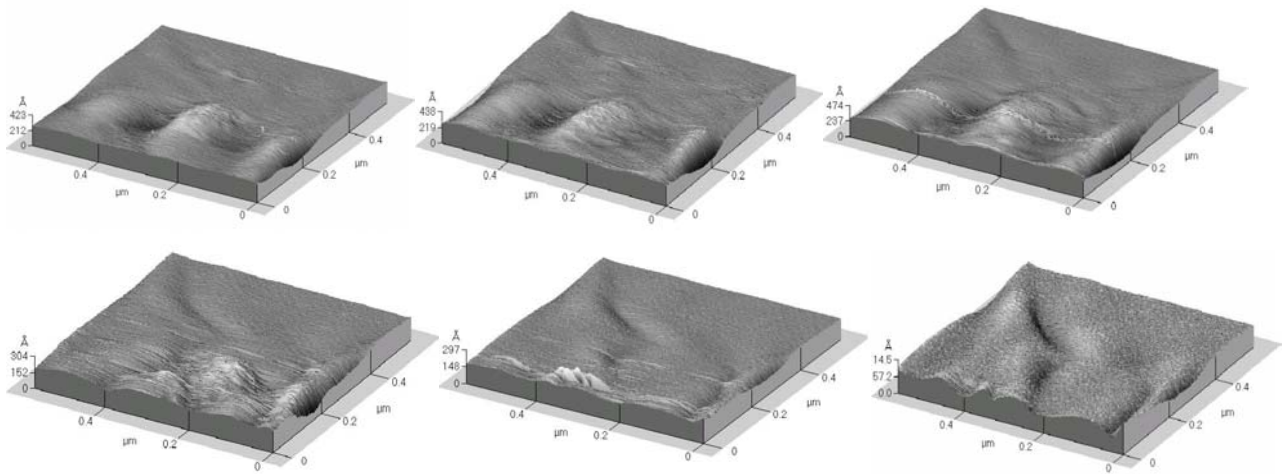

Figure 2. Sequence of six images of PMMA on the same area scratched by probe tip. The effect of plasticization of an asperity and formation of a ripple are clearly readable from the sequence.

The second step of the wear characterization regarded the qualitative analysis of polymer mechanical degradation. To highlight this process and possibly quantify it, a zero wear reference is need. The zero wear reference was formed covering partially an area of $SiO_2$ with the polymer and then to abrade that area at low loads and for a limited number of cycles (maximum of 15 scanning cycles). Under these experimental conditions, only the polymer would be exposed to wear due to the significantly higher hardness of $SiO_2$.

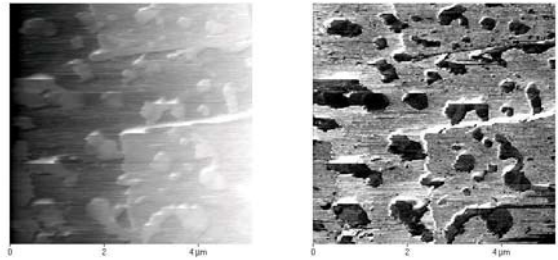

Figure 3. Surface of *SiO₂* covered by PMMA polymer. The left image shows the topography and the right, the correspondent FMM characterization. The black domains represent the islands of polymer.

In order to highlight the sample area covered by the polymer, mechanical characterisation has been performed making use of the Force Modulation Microscopy (FMM). FMM was used to detect variations in the mechanical properties of a surface, such as hardness. FMM allows simultaneous acquisition of both topography and material-properties data and it gives the possibility to distinguish the local areas where the polymer covers silicon oxide. In figure 3, the topography and the correspondent FMM image are shown, it is evident the separation between the area covered by the polymer and *SiO₂* substrate.

Figure 4 represents four different images acquired in different scanning cycles of the same area for PLLA (left), and PMMA (right). Image data are square matrices of 256×256 data points containing the height information of the measured surface area. Generally, AFM images of the same area acquired in sequence present lateral and rotational shifts. To be able to subtract accurately two consecutive images, they have to be aligned almost in x-y dimensions, before the amount of wear can be calculated. The alignment can be carried out making use of a numerical code, **N**anoscale **IM**age **A**lignment **C**ode (NIMAC) written *ad hoc*. In the numerical procedure the images should be adjusted following the steps: first, a frame compose by a matrix of pixels is defined for an image. Than, this frame is searched in a subsequent image making use of the statistical correlation. The application of these two steps was justified by the fact that all observed image shifts comprise translation movements and partially rotational [6,7]. After undergoing the adjustment procedures, the new surface images were compared quantitatively and the degree of change in wear volume was calculated.

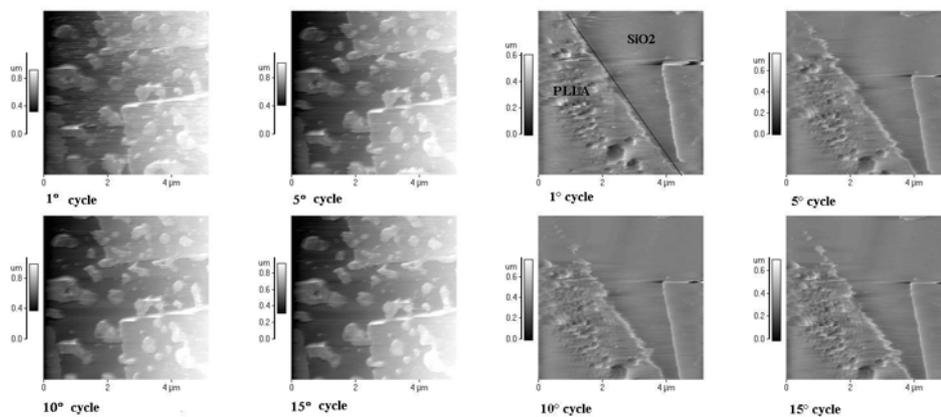

Figure 4. Four image series for PMMA and PLLA (applied load 1nN, pyramidal shape tip with nominal radius 40nm)

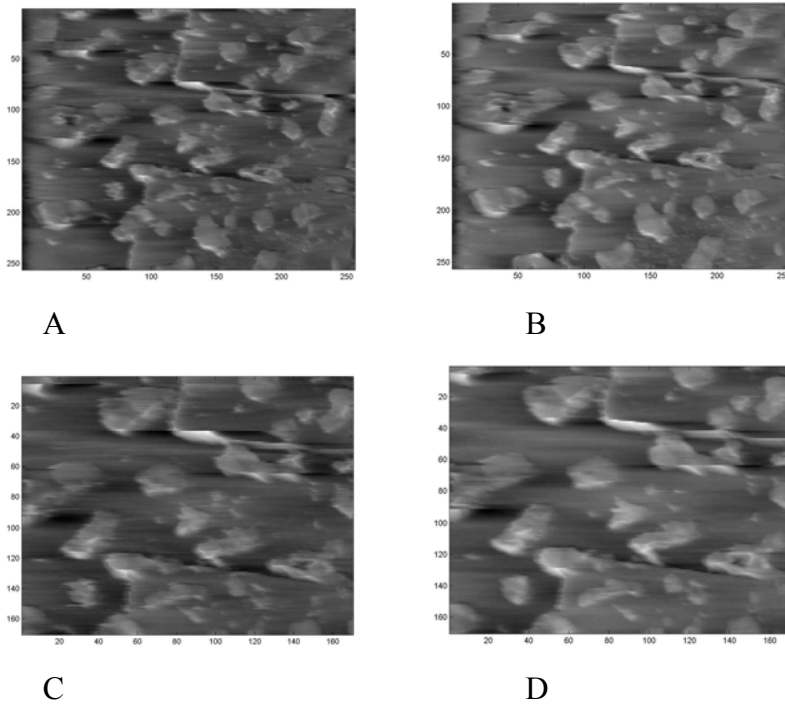

|     |     |
| :-: | :-: |
|  A  |  B  |
|  C  |  D  |

Figure 5. Application of NIMAC1.1 code for alignment of images. Figure 5.A shows the fifth scanning cycle and figure, 5.B shows the fifteenth scanning cycle. Figures 5.C-D represent the sub-area aligned of figures 5.A-B respectively.

Analogously to the procedure normally used in macroscale evaluation, the quantitative analysis of wear mechanisms could be carried out using volume change evaluation [13-15]. The wear volume can be calculated by integrating over the difference of the image data of the total area following the expression

$$\Delta V = V_B - V_A \qquad (2)$$

where $V_A$ and $V_B$ are two subsequent acquired AFM images after adjustment. In figure 6, the percentage of variation of volume is shown as a function of scanning cycles. Tribological process was rather complex, involving abrasion, adhesion, plastic deformation and variation of roughness. The behaviors of the volume changes $\Delta V$ as function of number of cycles show results difficult to be understood in straight manner. Generally, it could be expected that the wear volume tends to decrease as the number of stressing cycles increases due to plastic deformations of polymer. Nevertheless, the possible formation of ripples, as described above, should modify the polymeric surface making the volume changes due to abrasion difficult to observe. Resolution limits and errors in volume calculation could be given by tip geometry, thermal drift and undetected cantilever twist. The first two types of contributions should not be relevant if conical shape probe tip are used and the laboratory is thermally stable. On contrary, undetected cantilever twist could be a source of unexpectedly large errors. Other phenomena like electrical or mechanical noise should not contribute significantly to the error in volume estimation, since volume calculation is an averaging process and the net contribution should therefore be close to zero.

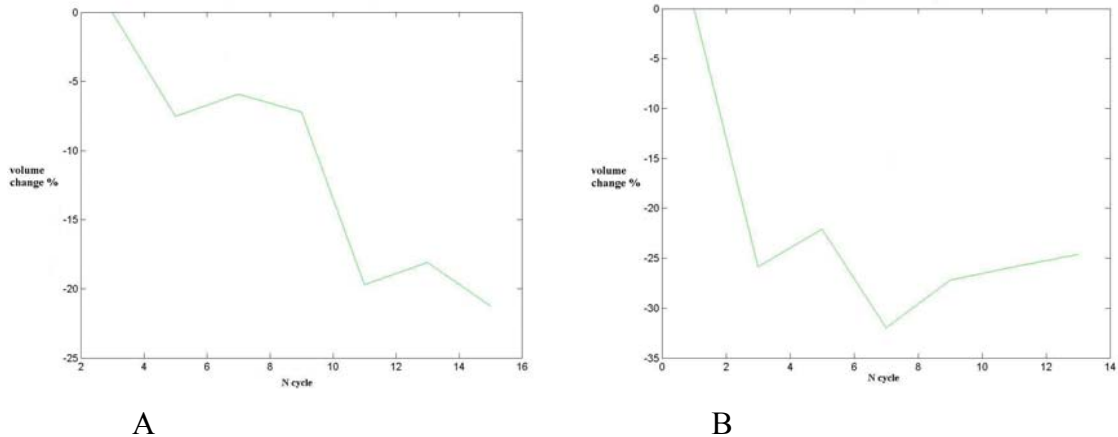

A                                                          B

Figure 6. Volume changes (percentage) for PMMA (A) and PLLA (B), respectively.

**Adhesion and tip wear.**
The third experiment has been conducted to highlight the crucial question of tip wear. The amount of tip wear can be obtained by observing the change of tip geometry before and after the sliding tests. Tip degradation can be generated by two main wear mechanisms, abrasion and adhesion wear-mechanisms. The abrasion wear-mechanism of tip could be activated by the sliding contact on silicon oxide, while adhesion wear-mechanism involves attachment of the polymer to the probe tip. SEM analysis of the tip demonstrated that the wear tests performed at 1nN did not expose the tip to significant wear mechanisms. This result confirmed that the experimental conditions were below the critical values of applied load and number of scanning cycles at which tip wear is observed. Nevertheless, the value for the wear depth/rate could be underestimated, because the nominal applied load (1nN) does not consider a supplementary effect due to adhesive interaction between AFM tip and polymers, being made the tests in air. The adhesive interaction should be stimulated by the presence of relative humidity. Adhesion force is the sum of capillary force, due to the Laplace pressure of the water meniscus forming at tip-sample interfaces and the direct adhesion of two contacting solids within the liquid, the adhesion force should involve a supplementary force of 0.1÷0.2nN [16,17].

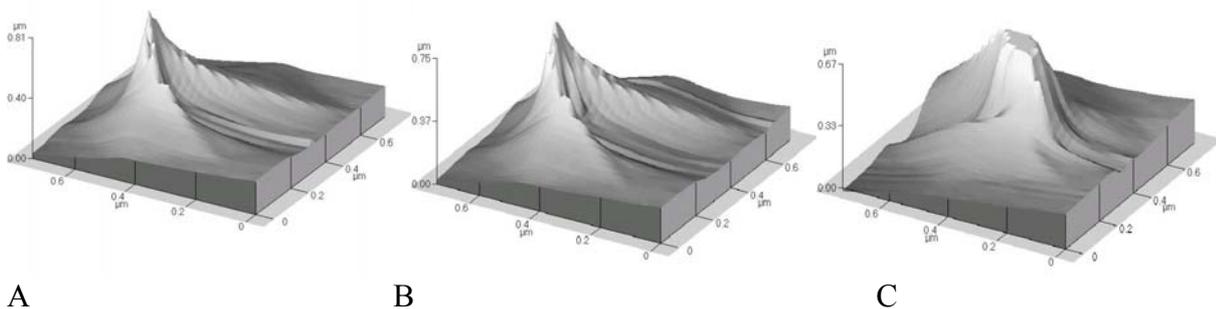

A                                B                                C

Figure 7. Images of tip degradation due to adhesion tip wear. Figure 7.A shows the convolution of a conical shape probe tip with the tip of grating tip, before the wear test. Figure 7.B shows the same probe tip after the wear test on PMMA. Figure 7.C shows the probe tip after the wear test on PLLA.

Generally, force-distance curves can help in order to provide a suitable analysis of tip wearing due to adhesion of polymer to the AFM tip [18]. The experiment to test the adhesion wear contribution was conducted first deposing the polymer on silicon oxide substrate, then rinsing the polymer with chloroform and finally scanning the polymer surface with a probe tip at 3nN. Drastic topographic changes were observed on PLLA. Imaging of δ-Silicon wafer grating was performed to evidence the tip degradation. Figure 7 shows the effect of material adhering to tip. The resultant image is the convolution of the probe tip and the grating tip. Probe tip degradation should give a deformed image of the grating tip. It is evident the polymer contamination of probe tip in the case of wear test on PLLA (figure 7.C).

**CONCLUSIONS**
Starting from these preliminary results, it has been shown that the presented surface analysis method using AFM is a valuable tool to study polymeric wear on micro/nanoscale. In particular, fundamental wear local phenomena located to a single asperity can be interpreted even when some environmental variables are not known accurately. The method could be used as a screening for study the wear resistance of bio-compatible polymers for use in biomedical applications. The methodology developed made it possible to observe and analyse qualitatively and, when possible, quantitatively some wear properties such as: i) formation of ripples on the surface of the polymers; ii) qualitatively evolution of the surface before and after testing; iii) evaluation of wear volume; iv) study of adhesion effect and subsequent degradation of the probe tip. In order to be able to distinguish between the simple wear of the polymer and the artifacts introduced by AFM tip wear rather than the convolution of sample wear and worn tip, precise alignments between subsequent scanning cycles has been reached as well as possible. The method described in this paper could be applied in precise environmental conditions such as polymer samples soaked in physiological buffers with determined pH.
The experimental and numerical methods adopted in this paper gave the possibility to have good indications on the incidence of some different wear-mechanisms. This made it possible to compare with a good degree of precision the wear resistance for different polymers employed in biomedical applications.

**Acknowledgements**
One author, M. D., would like to thank G. D. Guerra, CNR Pisa, S. Danti, L.P. Serino, University of Pisa, P. Pingue, Scuola Normale Superiore Pisa, and F. Vasca, University of Sannio, for helpful discussions.